\documentclass[fleqn,12pt,twoside]{article}
\usepackage{epsf,espcrc1}
\usepackage{graphicx}
\usepackage[figuresright]{rotating}

\newcommand{\AmS}{{\protect\the\textfont2
  A\kern-.1667em\lower.5ex\hbox{M}\kern-.125emS}}
\newcommand{\beq}{\begin{equation}}
\newcommand{\eeq}{\end{equation}}
\newcommand{\beqar}{\begin{eqnarray}}
\newcommand{\eeqar}{\end{eqnarray}}
\newcommand{\ds}{\displaystyle}
\title{Microscopic description of anisotropic flow in relativistic
heavy ion collisions}

\author{
E.~Zabrodin$^{a,c,d}$, L.~Bravina$^{b,d}$, C.~Fuchs$^a$, A.~Faessler$^a$ 
\\
\vspace*{.25 cm}
{\small\it
$^a$Institute for Theoretical Physics, University of
T\"ubingen, D-72076 T\"ubingen, Germany}\\
{\small\it
$^b$Department of Physics, University of Oslo, 
N-0316 Oslo, Norway}\\
{\small\it
$^c$Centre of Mathematics for Applications, University of Oslo, 
N-0316 Oslo, Norway}\\
{\small\it
$^d$Institute for Nuclear Physics, Moscow State University,
RU-119899 Moscow, Russia} \\
}
       
\begin{document}

\maketitle

\begin{abstract}
{\small
{\bf Abstract.}
Anisotropic flow of hadrons is studied in heavy ion collisions at
SPS and RHIC energies within the microscopic quark-gluon string
model. 
The model was found to reproduce correctly many of the flow
features, e.g., the wiggle structure of direct flow of nucleons 
at midrapidity, or centrality, rapidity, and transverse 
momentum dependences of elliptic flow. Further predictions are made.
The differences in the development of the anisotropic flow components 
are linked to the freeze-out conditions, which are quite different 
for baryons and mesons.
}
\end{abstract}


\section{INTRODUCTION}
\label{sec1}

The study of properties of extremely hot and dense nuclear matter, and 
the search for anticipated transition to a deconfined phase of quarks 
and gluons, the so-called Quark-Gluon Plasma (QGP), is one of the main 
objectives of heavy ion experiments at ultrarelativistic energies. 
Both theorists and experimentalists are looking for genuine 
QGP fingerprints, that cannot be masked or washed out by processes on 
a hadronic level. 
At present, the expansion of highly compressed nuclear mater in the
direction perpendicular to the beam axis of the colliding heavy ions,
known as collective flow, is believed to be one of the most promising
signals to detect the creation of the QGP \cite{QM02,ReRi97,HWW99}.
Since the development of flow is closely related to the equation of
state (EOS) of nuclear matter, the investigation of the flow can shed
light on the transition to the QGP phase accompanied by its 
subsequent hadronization
\cite{Amprl91,Olli92,HuSh95,RiGy96,Brprc94,Br95,Sorprl97,
HeLe99,CsRo99,Brac00,KSH00,ell_flow,TLS01,MCS02,LK02,MV03}.
If the transition from the QGP to hadronic phase is of first order,
the vanishing of the pressure gradients in the mixed phase leads to
the so-called softening of the EOS \cite{HuSh95,RiGy96}. The latter
should be distinctly seen in the behavior of the excitation function
of the collective flow.
This circumstance explains the great interest in the transverse flow 
phenomenon.

The Fourier expansion technique is usually employed to study 
collective flow phenomena since \cite{VoZh96,PoVo98}. 
The invariant distribution $ E d^3 N / d^3 p$ is presented as
\beq
\ds
E \frac{d^3 N}{d^3 p} = \frac{1}{\pi} \frac{d^2 N}{dp_t^2 dy} \left[
1 + 2 \sum_{n=1}^{\infty} v_n \cos(n\phi) \right] ,
\label{eq1}
\eeq
where $p_t$ and $y$ are the transverse momentum and the rapidity, and
$\phi$ is the azimuthal angle between the momentum of the particle
and the reaction plane. The first two Fourier coefficients in
Eq.~(\ref{eq1}), $v_1$ and $v_2$, are dubbed directed flow and
elliptic flow, respectively. Since both types of anisotropic flow
depend on rapidity $y$, transverse momentum $p_t$, and the impact
parameter of an event $b$ (i.e., $v_n \equiv v_n(x_j)$, where 
$\{ x_{j=1,2,3} \} \equiv \{ y, p_t, b \}$), the following 
differential distributions are usually applied
\beq \ds
v_n(x_i, \Delta x_{j \neq i}) =
\int_{x_j^{(1)}}^{x_j^{(2)}} \cos(n\phi) \frac{d^3 N}{d^3 x_j}
d^2 x_{j \neq i} \left/ \int_{x_j^{(1)}}^{x_j^{(2)}}
\frac{d^3 N}{d^3 x_j} d^2 x_{j \neq i} \right. \ .
\label{eq2}
\eeq

Model calculations suggest that elliptic flow is built up at the
early phase of nuclear collisions \cite{Sorprl97,KSH00,ell_flow}, 
whereas directed flow develops until the late stage of the reaction
\cite{LPX99,DF_prc00,DF_prc01}. 
But it is well known that the particles with high
transverse momentum are emitted at the onset of the collective
expansion, i.e., their directed flow can carry information about
the EOS of the dense nuclear phase. The study of the collective
flow development is, therefore, closely connected to the 
freeze-out picture.
In this article the microscopic quark-gluon string model (QGSM) 
\cite{qgsm1,qgsm2} is employed to investigate the formation and 
evolution of anisotropic flow components in heavy ion collisions at 
SPS ($\sqrt{s}=17.8$ AGeV) and RHIC ($\sqrt{s}=130$ and
200 AGeV) energies. Note that QGSM does not implement the formation of 
a QGP at the early stage of the collision. Our goal is to understand 
to what extent the characteristic signals of the hot nuclear matter 
can be reproduced. A noticeable discrepancy between experimental 
data and QGSM predictions being observed, this should be considered 
as an indication for new processes not included in the model.

\section{MODEL}
\label{sec2}

The QGSM is based on the $1/N_c$ (where $N_c$ is the number of quark
colors or flavors) topological expansion of the amplitude for
processes in quantum chromodynamics and string phenomenology of
particle production in inelastic binary collisions of hadrons. The
diagrams of various topology, which arose due to the $1/N_c$
expansion, correspond at high energies to processes with exchange of
Regge singularities in the $t$-channel. For instance, planar and
cylindrical diagrams corresponds to the Reggeon and Pomeron exchange,
respectively. The QGSM treats the elementary hadronic
interactions on the basis of the Gribov-Regge theory (GRT), similar to 
the dual parton model \cite{dpm} and the VENUS model \cite{venus}. 
This implies the consideration of subprocesses with quark 
annihilation and quark exchange, corresponding to Reggeon exchanges in 
two-particle amplitudes in the GRT, and with color exchange, 
corresponding to the one and more Pomeron exchanges in elastic 
amplitudes. The $hh$ collision term includes also single and double 
diffraction subprocesses, antibaryon-baryon annihilation and elastic
scattering, as well as the hard gluon-gluon scattering with large 
$Q^{2} > 1$ (GeV/$c$)$^{2}$ momentum transfer \cite{hard}. 

The inelastic {\it hh\/} cross section $\sigma_{in}(s)$
can be calculated via the real part of the eikonal 
\beq \ds
\sigma_{in}(s) = 2\pi \int \limits_{0}^{\infty}
\left\{ 1 - \exp\left[ - 2 u^{R}(s,b) \right]  \right\} b db \ .
\label{eq4}
\eeq
Here $s$ is the center-of-mass energy of the reaction.
The eikonal $u(s,b)$ can be presented as a sum of three terms
corresponding to soft and hard Pomeron exchange, and triple Pomeron
exchange, which is responsible for the single diffraction process,
\beq \ds
u^R(s,b) = u^R_{soft}(s,b) + u^R_{hard}(s,b) + u^R_{triple}(s,b)\ .
\label{eq5}
\eeq
Using the Abramovskii-Gribov-Kancheli (AGK) cutting rules \cite{agk}
the inelastic cross section of {\it hh\/} interaction can be presented
as
\beqar \ds
\sigma_{in}(s) &=& \sum \limits_{i,j,k = 0; i+j+k \geq 1}^{ }
\sigma_{ijk}(s)\ ,\\
\sigma_{ijk}(s) &=& 2 \pi \int \limits_{0}^{\infty} b db
\exp{\left[ -2 u^R(s,b) \right]} 
\frac{\left[ 2u^R_{soft}(s,b)  \right] ^i}{i !}
\frac{\left[ 2u^R_{hard}(s,b)  \right] ^j}{j !}
\frac{\left[ 2u^R_{triple}(s,b)  \right] ^k}{k !} \ .
\label{eq6-7}
\eeqar
The last equation enables one to determine the number of cut soft and
hard Pomerons, i.e., the number of strings and hard jets.
The single Pomeron exchange leads to the formation of two
quark-diquark or quark-antiquark strings. With rising energy
the processes with multi-Pomeron exchanges become
more and more important. The contribution of the cylinder diagrams
to the scattering amplitude increases like $s^{\alpha_P(0) -1}$, while
that of the so-called chain diagrams corresponding to $n$-Pomeron
exchanges $(n \geq 2)$ rises like $s^{n[\alpha_P(0) -1]}$ with
$\alpha_P(0) > 1$ being the intercept of a Pomeranchuk pole.
Strings, which are formed in the course of a {\it hh\/} or A+A
collision, decay later into secondary hadrons. Similar to hadronic
collisions, string fragmentation into hadrons proceeds independently
in A+A collision also. However, in the latter case these 
hadrons are allowed to interact with other hadrons after a certain 
formation time, while the valence quarks and diquarks can interact 
promptly with the reduced cross sections.
The model simplifies the nuclear effects and concentrates on hadron
rescattering. As independent degrees of freedom QGSM includes octet
and nonet vector and pseudoscalar mesons, and octet and decuplet
baryons, and their antiparticles.

The transverse motion of hadrons in the QGSM arises from different
sources: (i) primordial transverse momentum of the constituent quarks, 
(ii) transverse momentum of (di)quark-anti(di)quark pairs acquired at 
string breakup, (iii) the transverse Fermi motion of nucleons in
colliding nuclei, and (iv) rescattering of secondaries. Parameters of 
the first two sources are fixed by comparison with hadronic data.
The Fermi motion changes the effective transverse distribution of
strings formed by the valence quarks and diquarks of the target and
projectile nucleons. Thus, the original strings are not 
completely parallel to the beam axis. Further details of the QGSM
can be found elsewhere \cite{qgsm1,qgsm2}. 
We start from the study of energy and centrality dependence
of the directed flow.

\section{DIRECTED FLOW}
\label{sec3}

The directed flow of nucleons at BEVALAC/SIS energies
($E_{lab} = 0.1$ AGeV - 1 AGeV) and at AGS energy
($E_{lab} = 10.7$ AGeV) has a characteristic $S$-shape attributed to 
the standard $\langle p_x/A \rangle$ distribution. It grows linearly 
with rising rapidity between the target and projectile 
fragmentation regions. Conventionally, we will call this type of flow, 
for which the slope $d v_1/d y_{cm}$ is positive, {\it normal\/} flow, 
in contrast to the {\it antiflow\/} for which $d v_1/d y_{cm} < 0$ in 
the midrapidity region.

The one-fluid hydrodynamic models indicate that deviations from the
straight line behavior of the nucleon flow can be caused solely by the
softening of the EOS due to the QGP creation \cite{CsRo99}. In
microscopic string model calculations such deviations were first 
observed in very peripheral Au+Au collisions at AGS energy \cite{Br95}.
It appeared, however, that the effect is shifted to more central
topologies \cite{DF_prc00,DF_prc01} as the collision energy 
increases. The phenomenon leading to the formation of a characteristic
{\it wiggle\/} structure \cite{Sn00} of the directed flow is caused 
by shadowing, which plays a decisive role in the competition between
normal flow and antiflow in noncentral nuclear collisions at 
ultrarelativistic energies. 

\begin{figure}[htb]
\begin{minipage}[t]{75mm}
\vspace*{-2.0cm}
\centerline{\epsfysize=80mm\epsfbox{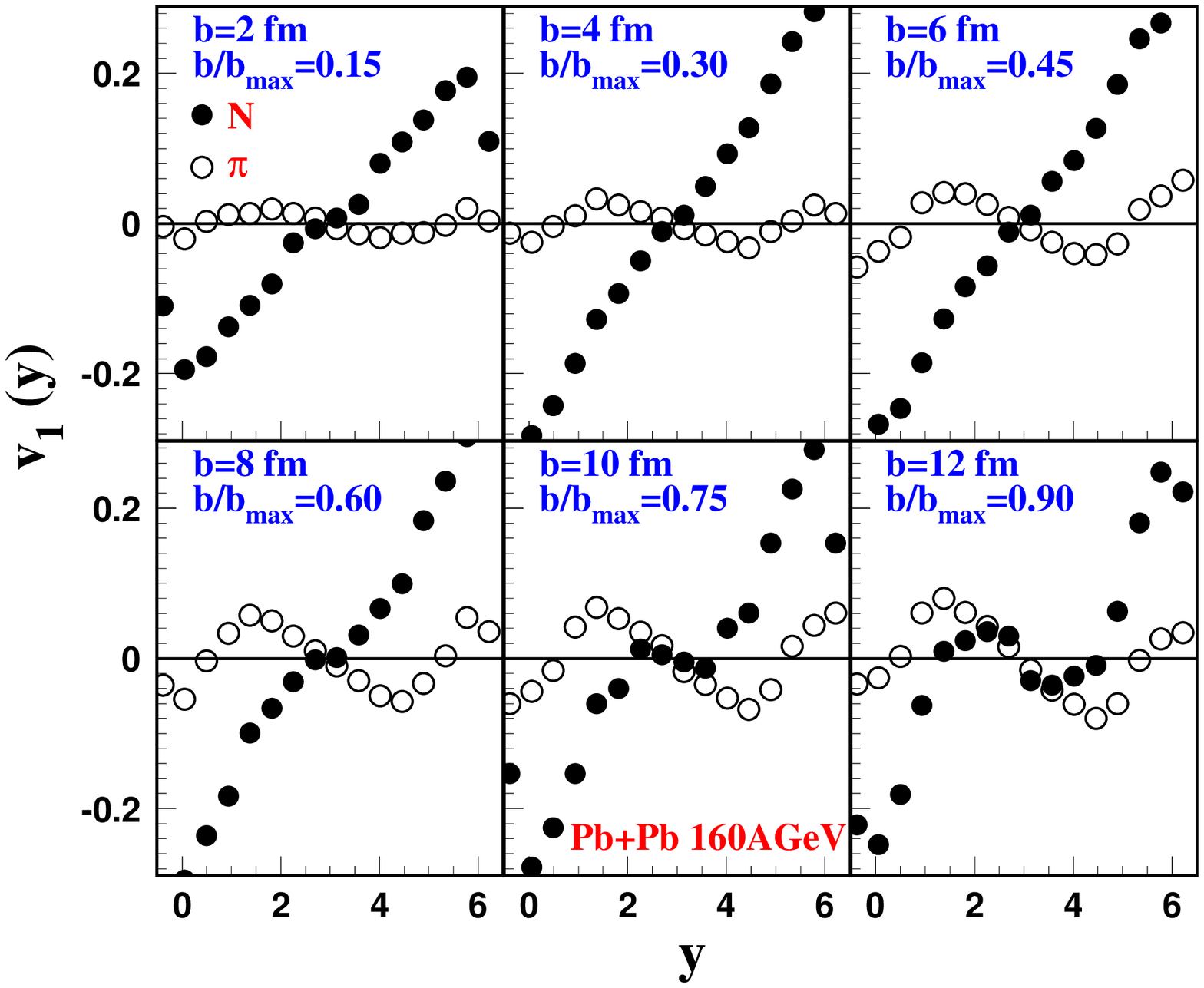}}
\vspace*{-1.5cm}
\caption{\small
Directed flow of nucleons (full circles) and pions (open circles) 
as a function of rapidity in lead-lead collisions at 160 AGeV.
 }
\label{fig1}
\end{minipage}
\hspace{\fill}
\begin{minipage}[t]{75mm}

\vspace*{-0.20cm}
\centerline{\epsfysize=50mm\epsfbox{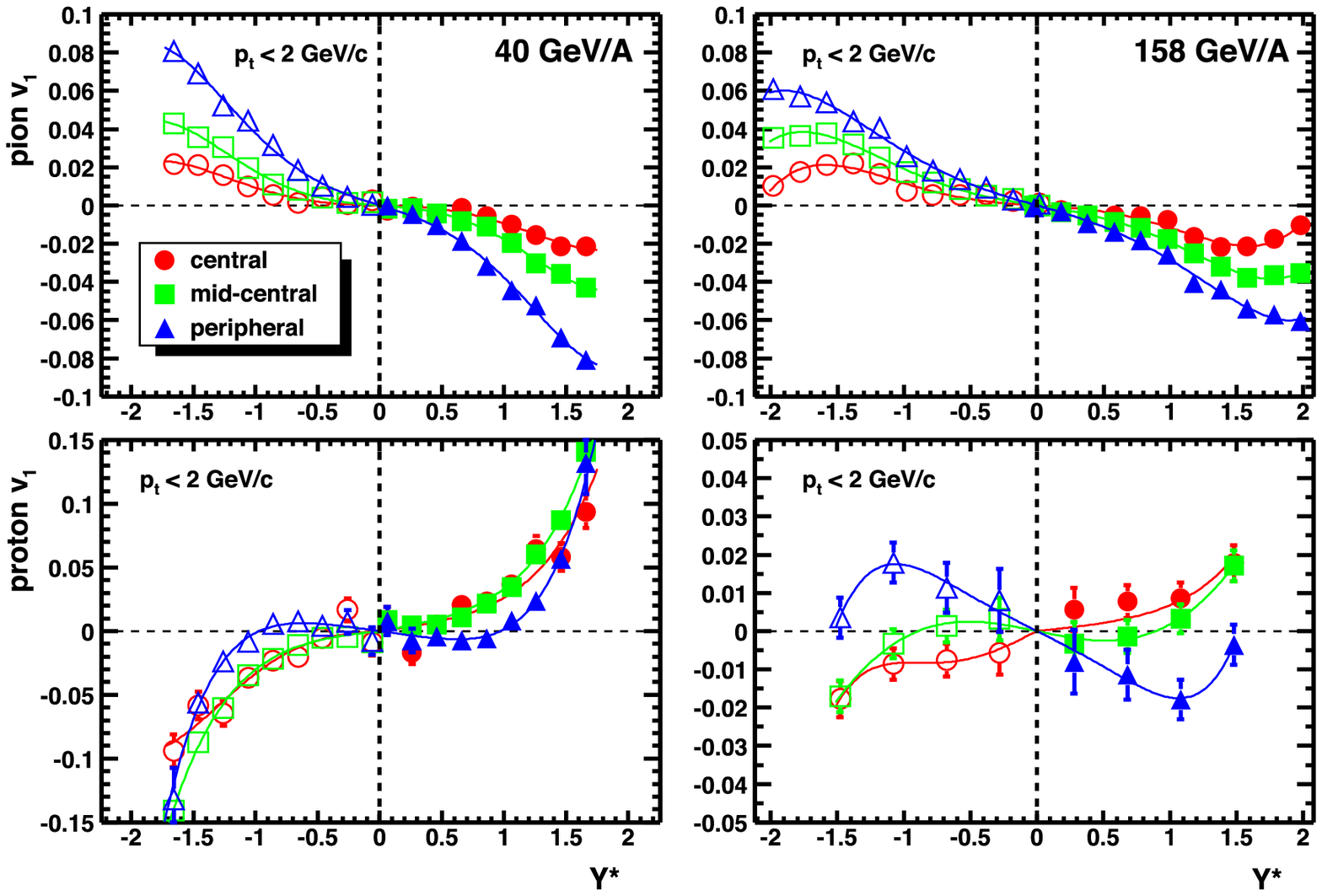}}
\vspace*{-0.30cm}
\caption{\small
Directed flow of protons and pions in Pb+Pb collisions at
40 AGeV and 160 AGeV measured by NA49 Collaboration
\cite{na49_wig}.
 }
\label{fig2}
\end{minipage}
\vspace*{-0.4cm}
\end{figure}
Hadrons, emitted with small rapidities in the antiflow area, can
propagate freely, while their counterparts will be absorbed by
dense baryon-rich matter. The rapidity dependence of the directed
flow of nucleons and pions in Pb+Pb at SPS is presented in
Fig.~\ref{fig1} for six different impact parameters. Here deviations
from the straight line start to develop at $b/b_{max} = 0.45$ in
the central rapidity window. Recently, the wiggle structure of
directed flow of protons in peripheral lead-lead collisions at
SPS has been observed by NA49 Collaboration \cite{na49_wig}
(see Fig.~\ref{fig2}). This result cannot be explained by the
effective softening of the EOS in microscopic models due to
production of strongly interacting string matter, because the  
number of the strings is significantly reduced as the collisions
become more peripheral. In hydrodynamics, even if the tilted region 
is non-perpendicular to the beam axis \cite{MCS02}, the tendency 
should be opposite: Here the antiflow reaches its maximum strength 
in semicentral events and almost disappear in peripheral ones.

Our next step is to study the time development of the directed flow of
strange hadrons, primarily, kaons and antikaons, in order to check
the role of baryon-antibaryon asymmetry.
For instance, $K^-$ and $\overline{K}^0$ can be absorbed via the
channels such as $K^- + p \rightarrow \Lambda + \pi^+$, etc.,
whereas there are no analogous reactions for $K^+$ and $K^0$.
How important is this reaction asymmetry at SPS and, especially, at
RHIC energies, where the matter is expected to be meson-dominated?
These problems have been addressed in \cite{kflow}.

\begin{figure}[htb]
\begin{minipage}[t]{75mm}
\vspace*{-0.7cm}
\centerline{\epsfysize=80mm\epsfbox{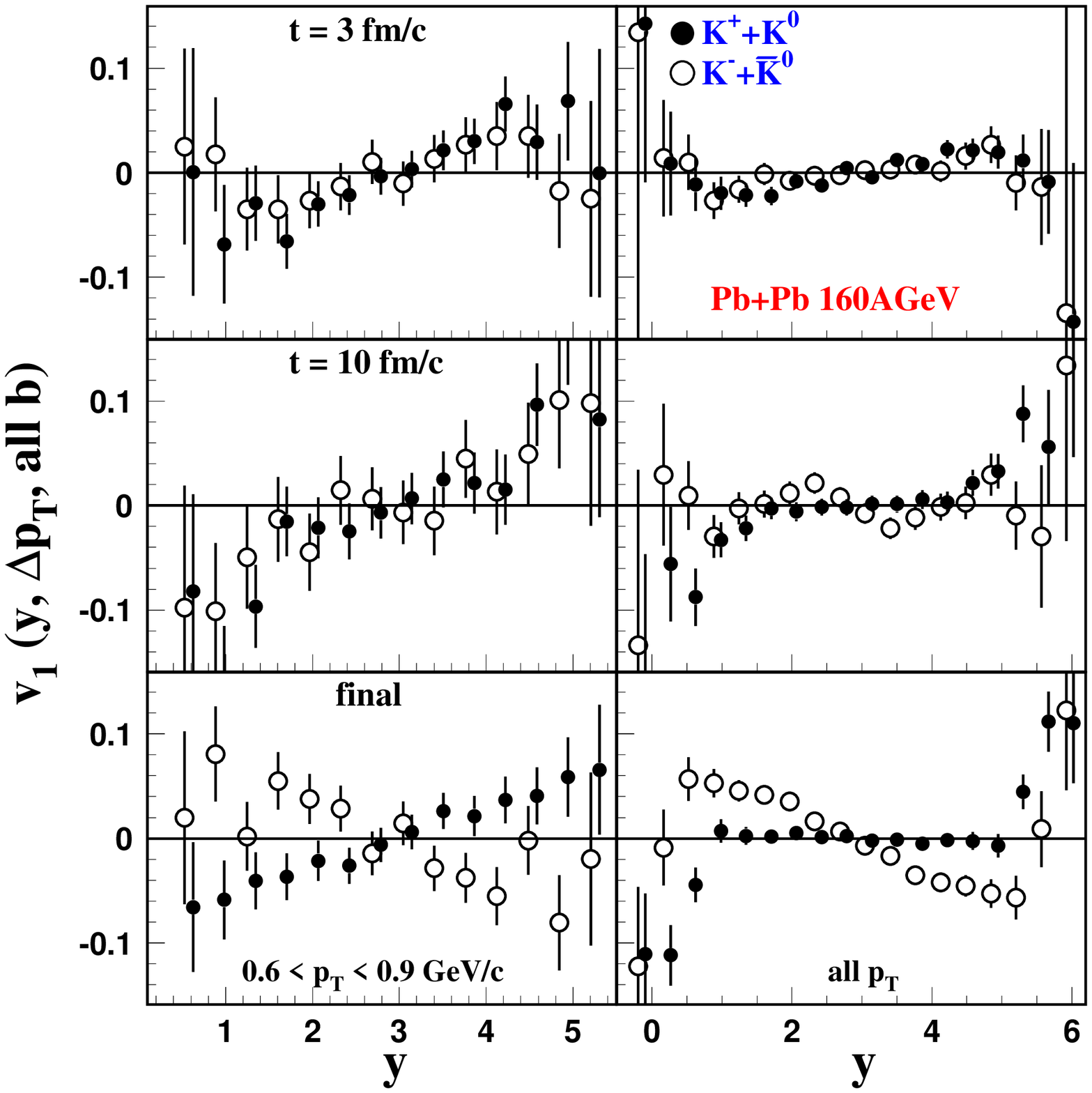}}
\vspace*{-1.0cm}
\caption{\small
$v_1^K(y)$ (solid circles) and $v_1^{\bar K}(y)$
(open circles) in min. bias Pb+Pb collisions at SPS
energy in high $p_t$ interval
$0.6 \leq p_t \leq 0.9$ GeV/$c$ (left) and for all $p_t$
(right panels) at times $t = 3$ fm/$c$, 10 fm/$c$, and final.
 }
\label{fig3}
\end{minipage}
\hspace{\fill}
\begin{minipage}[t]{75mm}

\vspace*{-0.70cm}
\centerline{\epsfysize=80mm\epsfbox{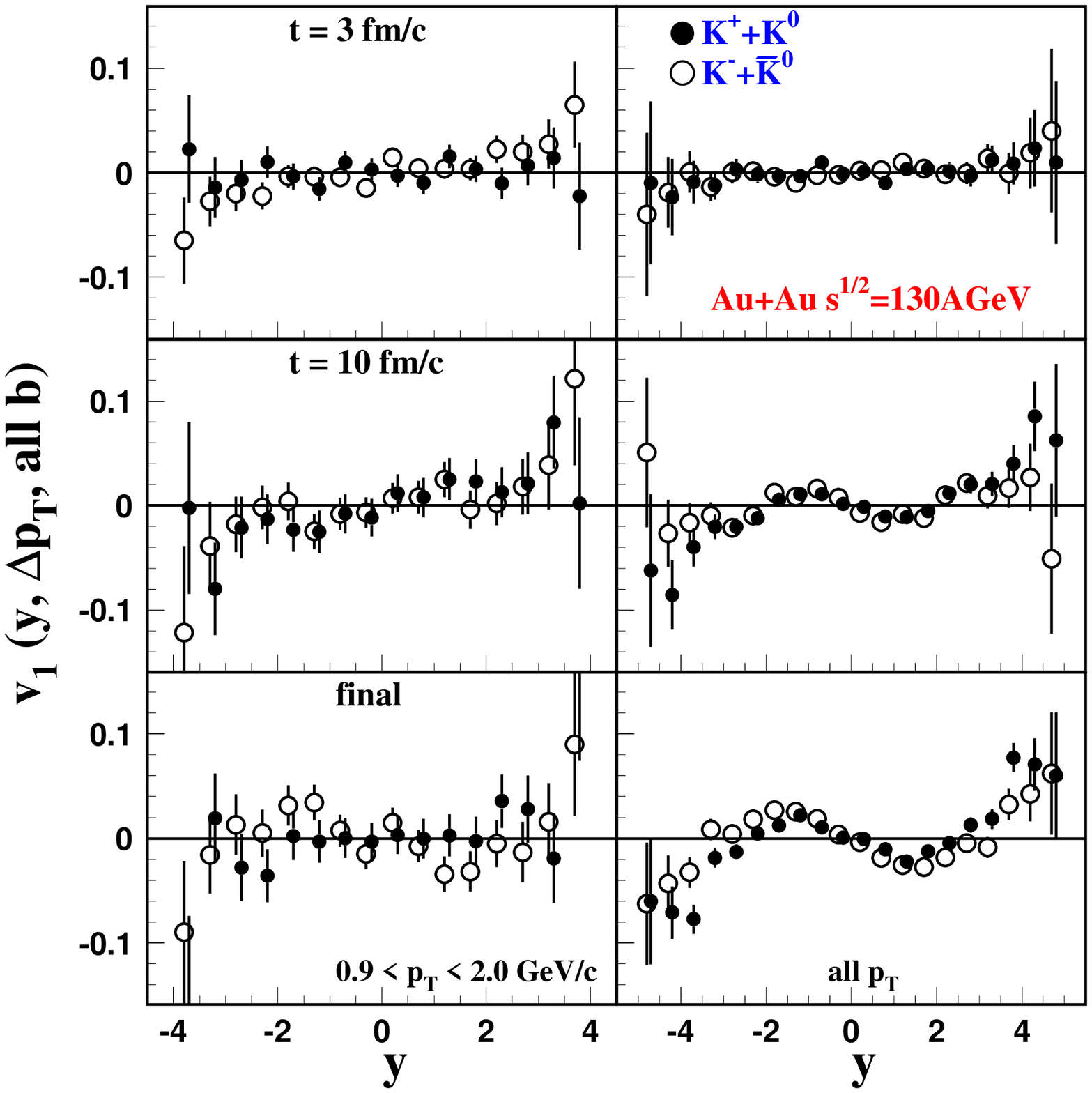}}
\vspace*{-1.00cm}
\caption{\small
The same as Fig.~3 but for minimum bias Au+Au collisions at RHIC
($\sqrt{s} = 130$ AGeV).
 }
\label{fig4}
\end{minipage}
\vspace*{-0.4cm}
\end{figure}

The time evolution of directed flow of kaons and antikaons in minimum
bias Pb+Pb collisions at 160 AGeV is presented in Fig.~\ref{fig3}.
Here the coefficient $v_1^K(y)$ is calculated in different
transverse momentum intervals at early, $t=3$ fm/$c$ and $t = 10$
fm/$c$, and the final stage of the reaction, $t \geq 60$ fm/c. To
avoid ambiguities, all resonances in the scenario with early
freeze-out were allowed to decay according to their branching ratios.
The flow evolution is seen quite distinctly. At early stages of the
collision directed flow of both kaons and antikaons is oriented in
the direction of normal flow similar to that of nucleons
\cite{DF_prc01}. Within the error bars there is no differences
between $(K^+ + K^0)$ and $(K^- + \overline{K}^0)$.
At this stage the matter is quite dense, mean free paths of particles
are short, and similarities in kaon production and rescattering
dominate over inequalities caused by different interaction
cross-sections. It is worth mentioning that the directed flow of $K$
and $\overline{K}$ is already sizable at $t = 3$ fm/$c$. This can be
explained by a kick-off effect associated with the early stage of the
collision, when the nuclei pass through each other. Later on the
system becomes more dilute. For both  kaons and antikaons the directed
flow experiences significant transformations. Already at $t = 10$ 
fm/$c$ the antiflow of
antikaons starts to built up in the midrapidity range. Note, that
$t = 10$ fm/$c$ corresponds to the maximum of the kaon $d N / d t$
distribution over their last elastic or inelastic interaction
\cite{FO_prc99}. Here the differences in interaction cross sections
and possible reaction mechanisms become crucial. 
Due to larger interaction cross-sections of $K^-$ and
$\overline{K}^0$ with other hadrons, the directed flow of these
particles changes the orientation from a weak normal to strong
antiflow. Even $(K^- + \overline{K}^0)$ with high transverse
momentum demonstrate distinct antiflow, while the flow of
$(K^+ + {K}^0)$ remains almost unchanged compared to that at
$t = 10$ fm/$c$.

The directed flow of kaons and antikaons in minimum bias Au+Au
collisions at $\sqrt{s} = 130$ AGeV is displayed in Fig.~\ref{fig4}
again at early
stages, $t = 3$ fm/$c$ and $t = 10$ fm/$c$, and at the final one,
$t \geq 100$ fm/$c$. It is interesting that at $t = 3$ fm/$c$ (i)
the flow of $(K^+ + K^0)$ coincides within the statistical
errors with the $(K^- + \overline{K}^0)$ flow, and (ii) the flow is
generally very similar to that at the SPS energy at time $t = 3$
fm/$c$. Except of the target and projectile fragmentation region,
where again the flow is probably produced by the initial kick, the
kaon flow at this early stage of gold-gold collisions at RHIC energy
is isotropic with respect to the impact parameter axis. The spatial
anisotropy in the distribution of baryonic charge seems to be
unimportant at this stage.  At $t = 10$ fm/$c$ not only the directed
flow of antikaons, but also that of kaons becomes antiflow-aligned at
midrapidity. Similar behavior has been found within the RQMD model
for the directed flow of nucleons at RHIC \cite{Sn00}, suggesting
that the nucleon directed flow is a side effect of the elliptic flow.
The flow of produced particles, pions \cite{Sn00,BlSt00} and kaons
\cite{BlSt00}, was found to be very flat at $|y| \leq 2$, in stark
contrast to the QGSM predictions. We are awaiting the experimental
data to resolve this problem.

\section{ELLIPTIC FLOW}
\label{sec4}

\begin{figure}[htb]
\begin{minipage}[t]{75mm}
\vspace*{-0.7cm}
\centerline{\epsfysize=80mm\epsfbox{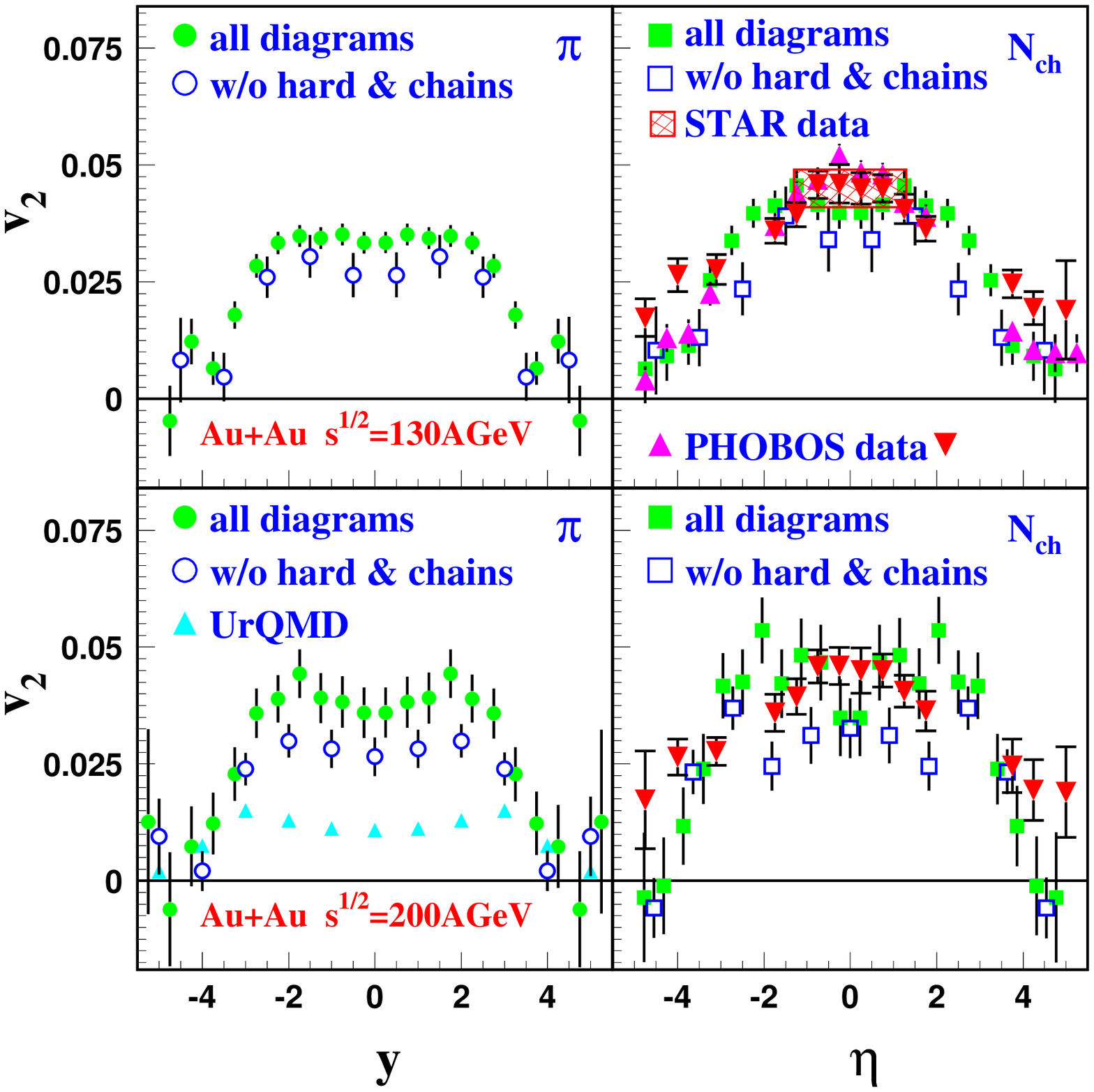}}
\vspace*{-1.0cm}
\caption{\small
Rapidity dependence of elliptic flow of pions and charged particles 
at RHIC energies. See text for details.
 }
\label{fig5}
\end{minipage}
\hspace{\fill}
\begin{minipage}[t]{75mm}

\vspace*{-1.40cm}
\centerline{\epsfysize=80mm\epsfbox{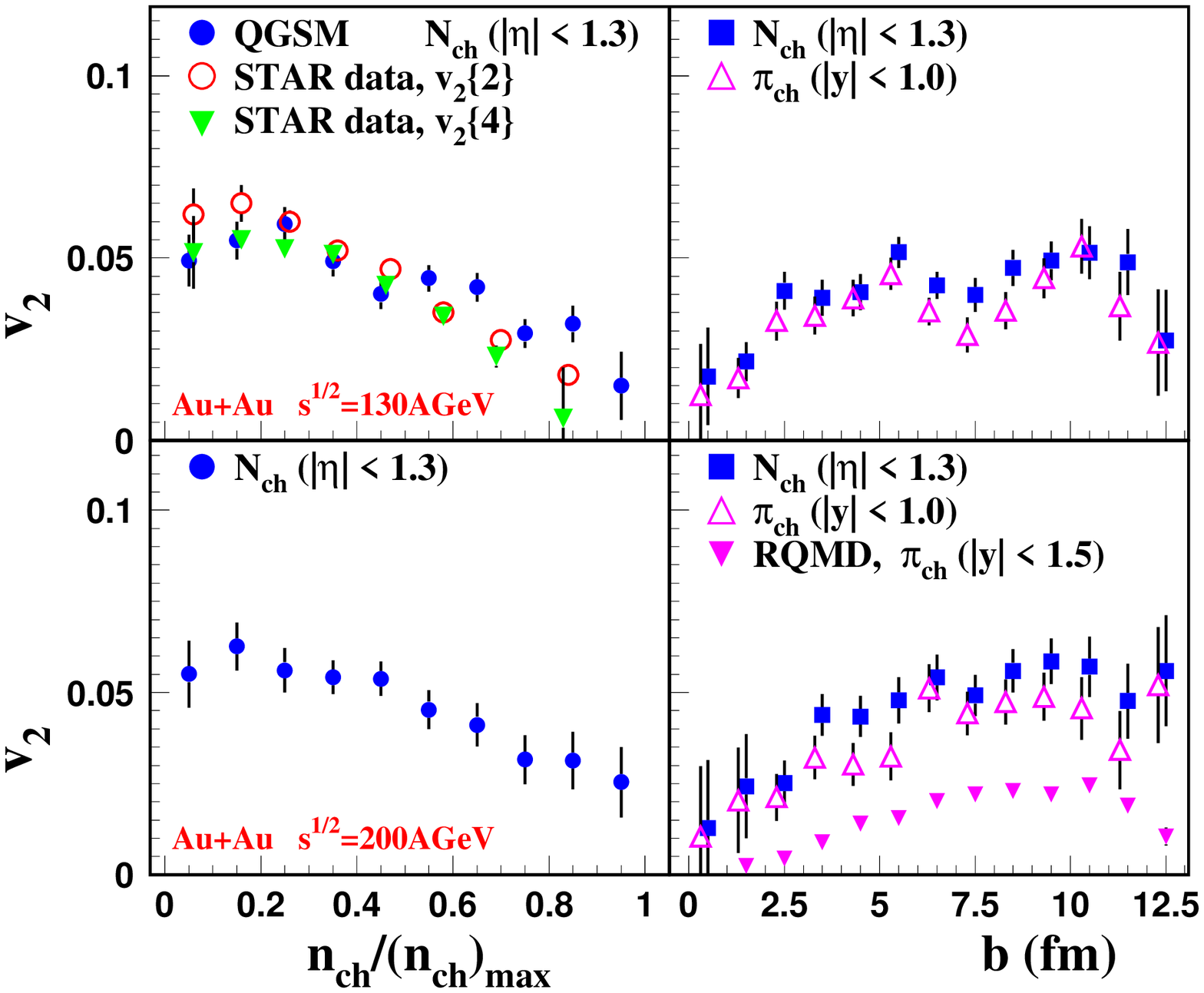}}
\vspace*{-0.40cm}
\caption{\small
Centrality dependence of elliptic flow of pions and charged particles
at RHIC energies.
 }
\label{fig6}
\end{minipage}
\vspace*{-0.4cm}
\end{figure}

Rapidity and pseudorapidity distributions of elliptic flow of
pions and charged particles at RHIC \cite{ell_flow} are depicted in 
Fig.~\ref{fig5}. For both energies elliptic flow displays strong 
in-plane alignment in accordance with the predictions of 
Ref.~\cite{Olli92}. In the mid(pseudo)rapidity the flow is almost 
constant. It rises slightly at $|y|, |\eta| \approx 1.7$, and then 
drops with increasing rapidity.
Pseudorapidity dependences of the elliptic flow of charged particles in 
the whole $\eta$ range, which were obtained {\it before\/} the
experimental data \cite{ell_flow}, are in a good agreement with the 
the results reported by PHOBOS Collaboration \cite{PHOB_fl}. 
To elaborate on the influence of hard processes and multi-Pomeron 
exchanges on the elliptic flow formation the flow caused by the 
subprocesses without the hard and multichain contributions is also 
plotted in Fig.~\ref{fig5}. It seems that in Au+Au collisions at
$\sqrt{s}=130${\it A} GeV the magnitude of the signal (except of the
midrapidity range) can be reproduced without the many-string
processes, although the particle multiplicity in the latter case is
reduced by 30\%. At $\sqrt{s}=200${\it A} GeV their role becomes more
significant, because the elliptic flow caused by other subprocesses
cannot exceed the limit of 3$-$3.5\% . Here it is important to
stress that the multichain diagrams alone, without rescattering,
cannot affect the elliptic flow at all. The flow increases solely
due to secondary interactions (absorption or rescattering) of produced 
particles in spatially asymmetric systems.

Figure \ref{fig6} presents the centrality dependence of elliptic
flow of charged particles. Since the centrality of events in the
experiment \cite{ellST00} has been determined via the ratio of 
charged particle multiplicity to its maximum value 
$N_{ch} / (N_{ch})^{max}$, we
compare the $v_2 \left[ N_{ch} / (N_{ch})^{max} \right]$ signal with
the original impact parameter dependence $v_2(b)$.
One can see that as a function of the impact
parameter $b$ elliptic flow is saturated at $b \approx 8$ fm for
both energies, while as a function of the multiplicity ratio
it increases nearly linearly with decreasing multiplicity up to
$N_{ch} / (N_{ch})^{max} \approx 0.15$. The agreement with the
STAR Collaboration data, obtained by 2-cumulant and 4-cumulant 
method \cite{ellST00,Sn03}, is good; similar results for the
centrality dependence of $v_2$ were obtained also by the PHENIX
Collaboration \cite{PHEN_fl}. As expected, the flow in
the midrapidity region is caused mainly by pions. The magnitude
of the pionic flow in the QGSM calculations is twice as large as
obtained, e.g. with RQMD \cite{SPV99}. Without many-string
processes the QGSM is able to describe the flow only in central and
semicentral collisions. It predicts a drop of the elliptic flow as
the reaction becomes more peripheral, which is similar to
the predictions of other string models.
Note that neither the pseudorapidity nor the centrality dependence
of the elliptic flow at RHIC is reproduced correctly so far by the
hydrodynamic models.

\begin{figure}[htb]
\begin{minipage}[t]{75mm}
\vspace*{-2.0cm}
\centerline{\epsfysize=77mm\epsfbox{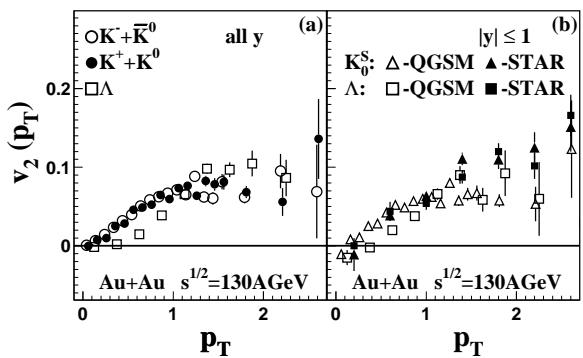}}
\vspace*{-2.0cm}
\caption{\small
Transverse momentum dependence of the elliptic flow of strange
particles in the whole rapidity (a) and in central rapidity interval
(b) in minimum bias Au+Au collisions at RHIC. Data taken from
\protect\cite{starKL}.
 }
\label{fig7}
\end{minipage}
\hspace{\fill}
\begin{minipage}[t]{75mm}

\vspace*{-0.8cm}
Figure~\ref{fig7} depicts the $p_T$-dependence of the elliptic flow
of strange hadrons at RHIC. The flow is close to zero for hadrons
with $p_T \leq 0.25$ GeV/$c$, then rises linearly (kaons) up to 
$v_2^K (p_T) \approx 10\%$ within the interval
$0.25 \leq p_T \leq 1.5$ GeV/$c$, and saturates at $p_T \geq
1.5\,$GeV/$c$ in accord with the experimental data \cite{starKL}.
The flow of lambdas is weaker than the kaon flow at $p_T \leq 1.3$
GeV/$c$. At higher transverse momenta the situation is changed: here
the $\Lambda$ flow is stronger. This is the general trend in the
$p_T$ dependences of mesonic and baryonic flow at RHIC energies.
The relative reduction of the elliptic flow within the interval
$1 \leq p_T \leq 2$ GeV/$c$ can be  explained by the interplay between 
the flow of high-$p_T$ particles, emitted at the onset 
\end{minipage}
\vspace*{-1.0cm}
\end{figure}
$\!\!\!\!\!$of the 
collision, and the hydro-type flow of particles, which gained their 
transverse momentum in subsequent secondary interactions.
Therefore, for proper understanding of the flow evolution it is
very important to investigate the freeze-out picture of the hadrons.

\section{FLOW AND FREEZE-OUT}
\label{sec5}

\begin{figure}[htb]
\begin{minipage}[t]{75mm}
\vspace*{-0.7cm}
\centerline{\epsfysize=80mm\epsfbox{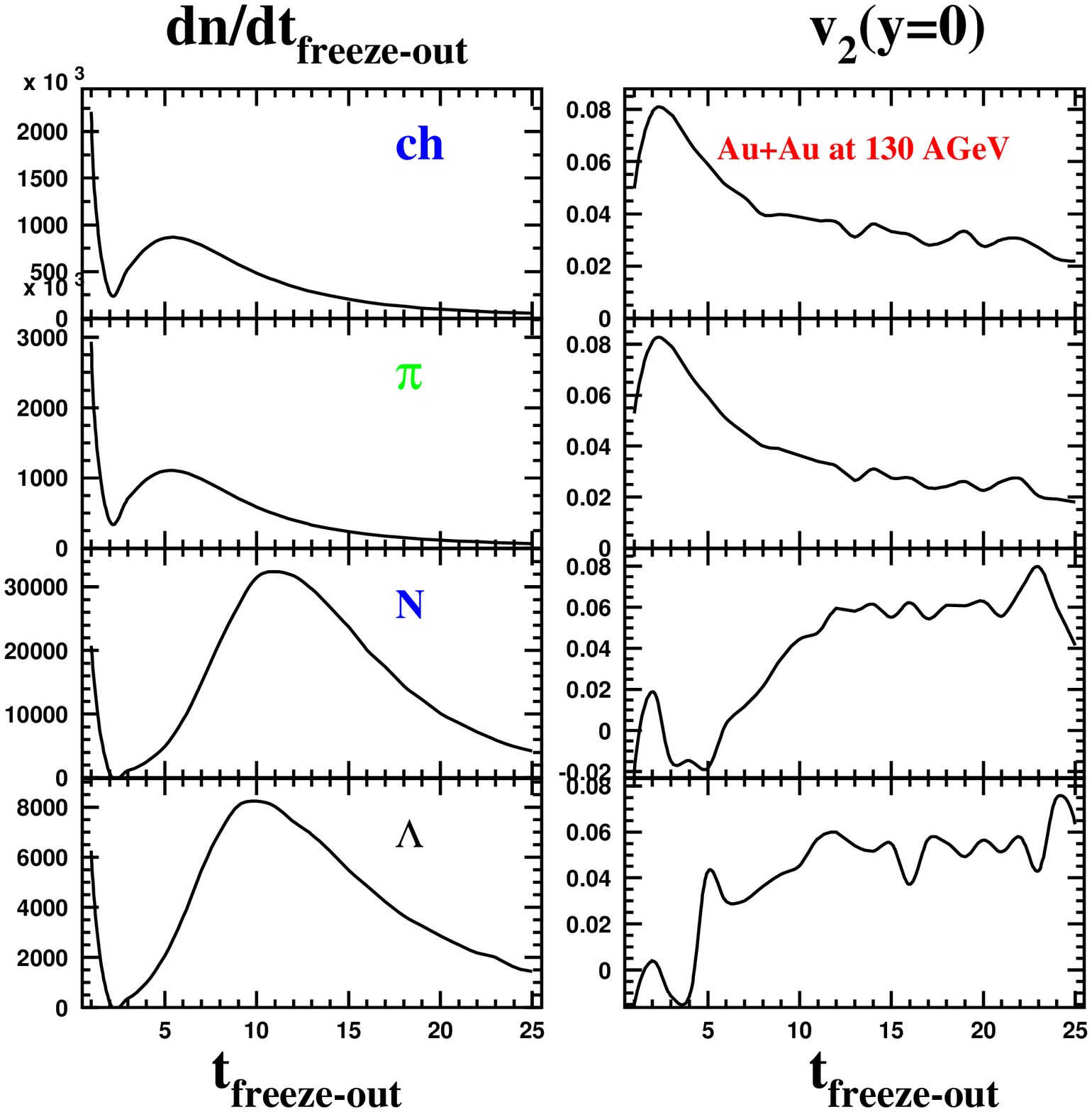}}
\vspace*{-0.6cm}
\caption{\small
$dN/dt$ distribution of $n_{ch}, \pi, N, \Lambda$ over
the time of their last interaction (left) and elliptic flow of
these particles (right) for Au+Au collisions with $b=8$ fm at
$\sqrt{s}=130$ AGeV.
 }
\label{fig8}
\end{minipage}
\hspace{\fill}
\begin{minipage}[t]{74mm}
\vspace*{-0.55cm}
\centerline{\epsfysize=77mm\epsfbox{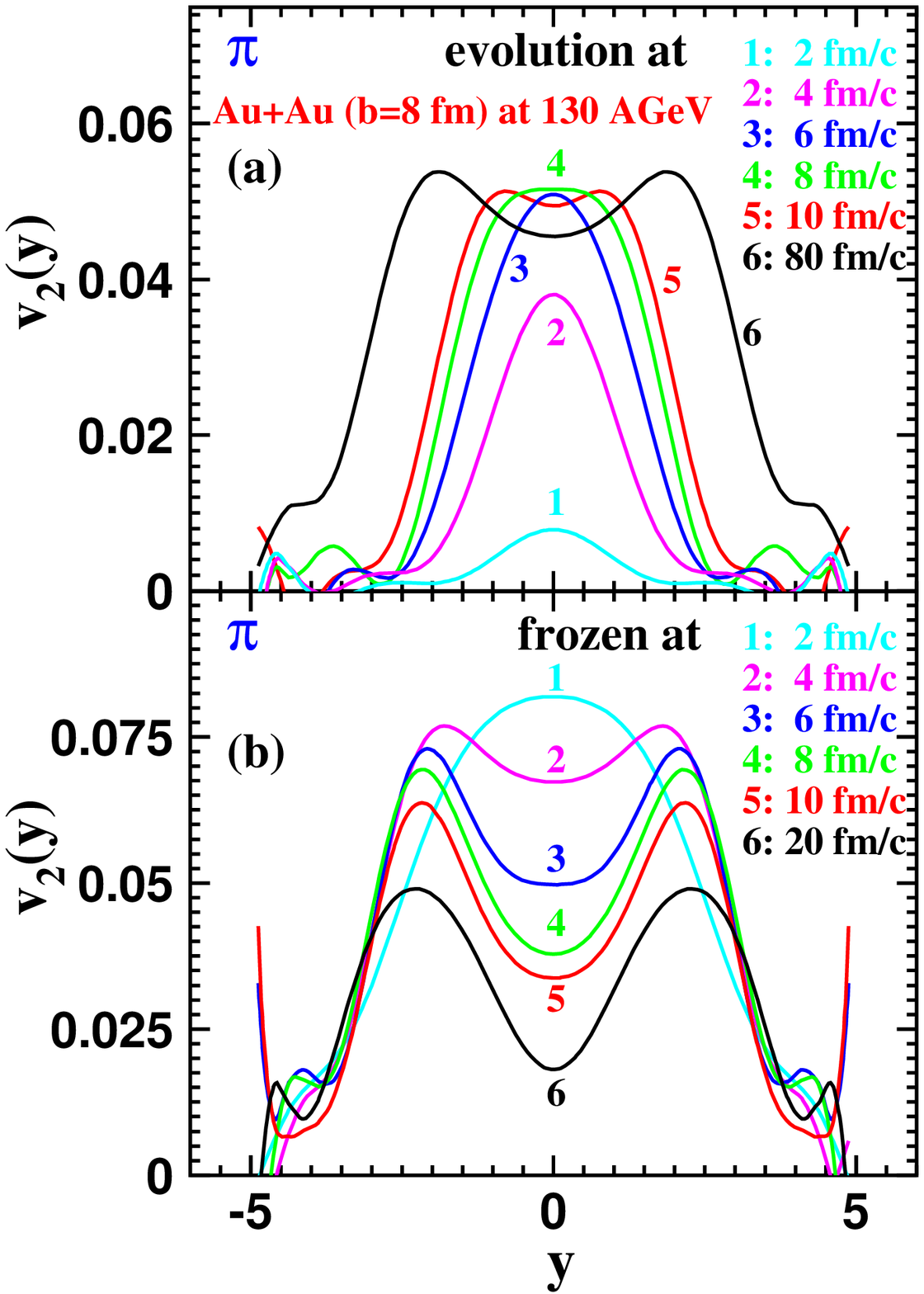}}
\vspace*{-0.5cm}
\caption{\small
Gold-gold collisions with $b=8$ fm at $\sqrt{s}=130$ AGeV:
(a) Snapshots of the $v_2^\pi$ at 2, 4, ... 80 fm/$c$ (from bottom to
top), and (b) elliptic flow of pions already frozen at 2, 4, ...
20 fm/$c$ (from top to bottom).
 }
\label{fig9}
\end{minipage}
\vspace{-0.5cm}
\end{figure}

The $dN/dt$ distributions of $n_{ch}, \pi, N, \Lambda$, which are 
decoupled from the fireball after the last elastic or inelastic 
collision, are shown in Fig.~\ref{fig8}. For the analysis ca. 20000
gold-gold collisions with the impact parameter $b=8$ fm at 
$\sqrt{s} = 130$ AGeV were simulated. One can see, that the 
particles are emitted during the whole course of the system evolution.
In this respect the freeze-out picture obtained in microscopic model
is different from the sharp freeze-out assumed in hydrodynamic models.
Compared to AGS and SPS 
energies \cite{FO_prc99}, a substantial part of hadrons leave the 
system immediately after their production within the first two fm/$c$. 
The pion distribution 
has a peak at $t = 5-6$ fm/$c$, while the distributions of 
baryons are wider due to the large number of rescatterings, which 
shift their $dN/dt$ maxima to later times, $t = 10 - 12$ fm/$c$.
Since the pion fraction dominates the other hadrons, the distribution
of charged particles is similar to the pion one.  

The elliptic flow carried by these species is presented in 
Fig.~\ref{fig8} as well. The baryonic and mesonic components are
completely different: pions emitted from the surface of the expanding
fireball within the first few fm/$c$ carry the strongest flow, while
later on the flow of pions is significantly reduced. In contrast to this, 
the baryon fraction acquires stronger elliptic flow during the
subsequent rescatterings, thus developing the hydro-like flow.
The saturation of the flow at the late stages can be explained
by the lack of rescattering, since the expanding system becomes
more dilute. 
 
Finally, the time evolution of the elliptic flow of pions at
midrapidity is studied. Two varieties of rapidity distribution are 
displayed in Fig.~\ref{fig9}: elliptic flow of pions frozen at
$t =2$ fm/$c$, 4 fm/$c$, etc. till the late stages of the reaction;
and the rapidity profile of the total pion elliptic flow at the same 
times. In the latter case, similarly to Sec.~\ref{sec3}, the 
interactions were switched off, but resonances were allowed to decay. 
Here two observations should be mentioned. Firstly, although the 
elliptic flow at $y = 0$ reaches the maximum already at 
$t \approx 6$ fm/$c$, its formation is not over due to continuous 
freeze-out of particles. Secondly, there is no one-to-one 
correspondence between the apparent elliptic flow and the contribution 
to the final flow coming from
the ``survived" fraction of particles. For instance, apparent elliptic
flow of pions at $t =2$ fm/$c$ is weak, but pions which are frozen at
this moment have the strongest elliptic anisotropy caused by the 
absorption of the pion component in the squeeze-out direction.

\section{CONCLUSIONS}
\label{sec6}

In summary, the features of directed flow in the model can be
stated as follows:\\
(i) wiggle structure of nucleon flow at midrapidity (SPS, RHIC);\\
(ii) the negative slope of the flow at $|y| \leq 2$ similar for all
hadrons (RHIC); \\
(iii) normal flow of high-$p_t$ hadrons at
midrapidity (SPS, RHIC); \\
(iv) flows of kaons and antikaons are
different at SPS but similar at RHIC.
 
Elliptic flow features at RHIC:\\ 
(i) flow at $\sqrt{s} = 200$ AGeV 
is only 10-15\% stronger than that at $\sqrt{s} = 130$ AGeV;\\
(ii) $v_2(\eta)$ varies slightly within the interval $|\eta| \leq 
2$ and quickly drops at $|\eta| > 2$;\\ 
(iii) $v_2(p_t)$ of mesons
is stronger than the baryonic flow at $p_t \leq 1.5$ GeV/$c$ but
saturates earlier;\\ 
(iv) the time evolution of the mesonic flow is opposite to that of 
the baryonic one.\\
Many of these peculiarities are linked to
different freeze-out pictures for baryons and mesons. In particular,
it is shown that the flows of particles emitted at the beginning
of the collision and during the course of the reaction are quite
different. Therefore, phenomena such as collective flow and
particle freeze-out should not be considered independently.  
The collective flow of hadrons appears to have a multi-component 
structure, caused by rescattering and absorption in a spatially
anisotropic medium, which deserves further investigations. 

{\small {\it Acknowledgments.}
Fruitful discussions with L. Csernai, R. Lacey, C.M. Ko, 
E. Shuryak, and N. Xu are gratefully acknowledged.
This work was supported by the Deutsche Forschungsgemeinschaft, the 
Bundesministerium f{\"u}r Bildung und Forschung under contract 
06T\"U986, and the Norwegian Research Council (NFR).
}


\end{document}